\documentclass[prb,aps,amsmath,amssymb,amsfonts,floatfix,superscriptaddress,showpacs,twocolumn,nofootinbib]{revtex4-1}
\usepackage{graphicx}
\usepackage{color}
\usepackage{bbm}
\usepackage[utf8]{inputenc}
\usepackage{dcolumn}
\usepackage{hyperref}
\usepackage[caption=false]{subfig}
\usepackage[english]{babel}
\usepackage{indentfirst}
\usepackage{textcomp}
\hypersetup{colorlinks=true,breaklinks,linkcolor=blue,urlcolor=blue,citecolor=blue}

\usepackage{lipsum, babel}

\newcommand{\qv}{\ensuremath{\mathbf{q}}}

\newcommand{\dv}{\ensuremath{\hat{d}}} 

\newcommand{\av}[1]{\ensuremath{\left\langle #1 \right\rangle}}

\def \half {\ensuremath{\frac{1}{2}}}

\begin{document}

\title{Ultralong-range order in the Fermi-Hubbard model with long-range interactions}

\author{Erik G. C. P. van Loon}
\affiliation{Radboud University, Institute for Molecules and Materials, NL-6525 AJ Nijmegen, The Netherlands}

\author{Mikhail I. Katsnelson}
\affiliation{Radboud University, Institute for Molecules and Materials, NL-6525 AJ Nijmegen, The Netherlands}

\author{Mikhail Lemeshko}
\email{mikhail.lemeshko@ist.ac.at}
\affiliation{IST Austria (Institute of Science and Technology Austria), Am Campus 1, 3400 Klosterneuburg, Austria} 

\begin{abstract}

We use the  dual boson approach to reveal the phase diagram of the Fermi-Hubbard model with long-range dipole-dipole interactions. By using a large-scale finite-temperature calculation on a $64 \times 64$ square lattice we demonstrate the existence of a novel phase, possessing an  `ultralong-range'  order. The fingerprint of this phase -- the density correlation function -- features a non-trivial behavior on a scale of tens of the lattice sites. We study the properties and the stability of the ultralong-range ordered phase, and show that it is accessible in modern experiments with ultracold polar molecules and magnetic atoms.

\end{abstract}


\maketitle

Recent experimental progress opened up a possibility to use ultracold quantum gases in optical lattices to realise exotic many-particle Hamiltonians inaccessible in `conventional' condensed matter physics~\cite{LewensteinBook12, BlochNatPhys12, BaranovCRev12}. 
One of the rapidly advancing research directions deals with particles possessing a dipole moment, such as magnetic atoms~\cite{GriesmaierPRL05, NaylorPRA15, LuPRA11, LuPRL12, AikawaPRL12, AikawaPRL14} or ground-state heteronuclear molecules~\cite{OspelkausFDiss09, NiNature10, OspelkausScience10, TakekoshiPRL14, MolonyPRL14, ParkPRL15, Park15b}. The anisotropic and long-range character of the dipole-dipole interactions between such species is predicted to give rise to novel phases of matter~\cite{BaranovCRev12, KreStwFrieColdMolecules, CarrNJP09, LemKreDoyKais13, LahayePfauRPP2009, BaranovPRep08, TrefzgerJPB11}. To date, several many-body models have already been implemented in the laboratory~\cite{dePazPRL13, YanNature13}.

In particular, optical lattice experiments allow to simulate the Dipolar Fermi-Hubbard (DFH) model, i.e.\ an extended Hubbard model with long-range dipole-dipole interactions~\cite{DuttaRPP15}. While the phase diagram of its bosonic counterpart has been evaluated using large-scale quantum Monte Carlo simulations~\cite{CapogrossoPRL10, PolletPRL10}, understanding the DFH model represents a formidable challenge due to the sign problem in quantum Monte Carlo~\cite{DeRaedt85,PolletRPP12}. 

Recently, the DFH model has been approached using a number of techniques, starting from traditional mean-field~\cite{MikelsonsPRA11, Gadsbolle12-2,Gadsbolle12} and Hartree-Fock-Bogoliubov approximations~\cite{ZengPRB14}. 
The Functional Renormalization Group technique~\cite{ShankarRMP94} has uncovered novel bond-ordered phases in systems of dipolar fermions at half-filling~\cite{Bhongale12,Bhongale13}. 

However, it is unclear how applicable it is when the local and dipole-dipole interaction strengths are comparable with the kinetic energy. 
Furthermore, most optical lattice experiments take place in a harmonic trap, with different kinds of fillings present simultaneously. Therefore, it is important to understand the behavior of the system away from the special case of half-filling.

Dynamical Mean Field Theory (DMFT)~\cite{Metzner89,Georges96} has been extensively used for electronic structure theory~\cite{Kotliar06}. It can be applied both at small and at large local interaction strength.
This method has been adapted to the study of optical lattices in the form of real-space DMFT~\cite{Snoek08,Helmes08} and used to study phase transitions in half-filled dipolar fermion systems~\cite{Hofstetter11}. However, the nonlocal interaction had to be restricted to its Hartree contribution. In real-space DMFT a single-site problem has to be solved for every lattice site, hindering the study of large lattice sizes.

\begin{figure}[b]
 \includegraphics[width=0.95\linewidth]{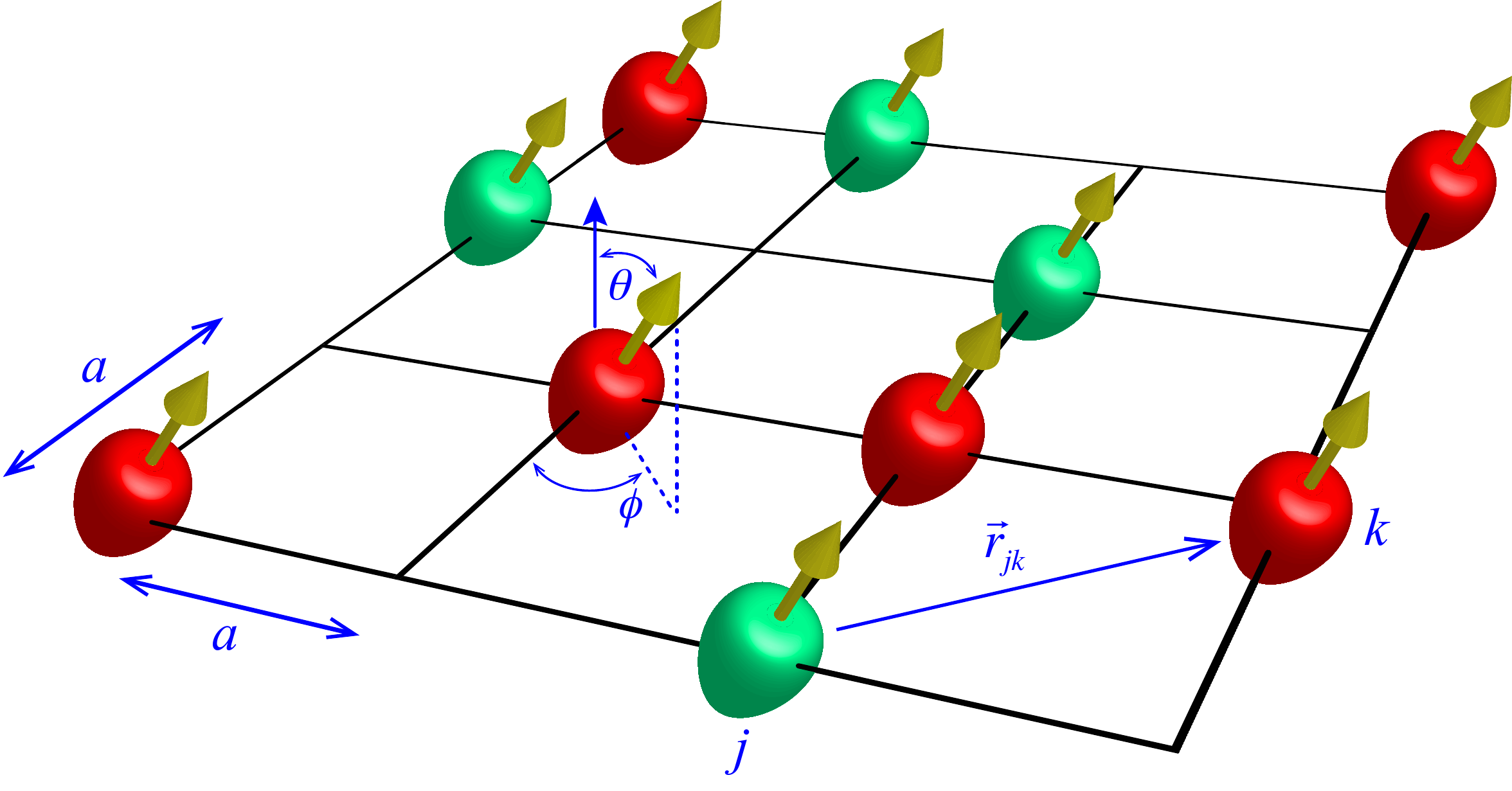}
 \caption{ (Color online) Dipolar fermions in a square optical lattice. The orientation of the dipoles is given in by the spherical angles $(\theta, \phi)$.}
 \label{fig:setup}
\end{figure}

Here, we reveal the phase diagram of the DFH model using the dual boson approach~\cite{Rubtsov12}. The method is based on a single-site impurity problem and therefore allows to treat lattices of larger sizes, which is crucial for systems featuring  long-range interactions. 
On the other hand, the technique is applicable even when the interaction strength is comparable to the kinetic energy. Furthermore, the technique allows to perform finite-temperature calculations away from half-filling, which is crucial in order to reproduce the conditions of realistic experiments. 
As the main result, our large-scale calculation allows to demonstrate  the occurrence of a novel phase, featuring `ultralong-range' density correlations at distances of tens of lattice sites. While such a phase has never been predicted before, it appears to be within the reach in modern experiments with dipolar quantum gases.

We start with a two-component gas of dipolar fermions trapped in a two-dimensional square optical lattice, as schematically illustrated in Fig.~\ref{fig:setup}, and described by the DFH Hamiltonian:
\begin{align}
\label{eq:hmlt}
  H = - t \sum_{\av{jk}\sigma} c_{j\sigma}^\dagger c_{k\sigma}^{\phantom{\dagger}} +  \frac{U}{2} \sum_{j} n_j n_j + \half \sum_{jk} V^d_{jk} n_j n_k
\end{align}
Here $c_{j\sigma}^\dagger$ ($c_{j\sigma}^{\phantom{\dagger}}$) is the creation (annihilation) operator for a fermion with spin state $\sigma$ on site $j$, and $\av{jk}$ is a pair of nearest neighbours. 
In experiment, the two spin components $\sigma$ can be represented by different fine, hyperfine, or rotational states of the dipolar species used. 
The first two terms of eq. (\ref{eq:hmlt}) give the amplitudes of the nearest-neighbor hopping, $t$, and the on-site interaction, $U$.  The third term corresponds to the dipole-dipole interaction, whose spatial dependence is given by $V^d_{jk} = c_d \left[1-3(\hat{r}_{jk}\cdot \dv)^2 \right]/(r_{jk}/a)^3$. Here $\mathbf{r}_{jk}$ is the vector connecting the fermions on sites $j$ and $k$, with $\hat{r}_{jk} = \mathbf{r}_{jk}/r_{jk}$, and $a$ is the lattice constant. The fermions' dipole moments point in the same direction given by the unit vector $\dv$. The dipole-dipole interaction strength parameter depends on the particular species involved and is given by $c_d = d^2/(4 \pi \epsilon_0)$ for the electric dipoles of magnitude $d$ in the laboratory frame, and by $c_d = \mu_0 \mu^2/(4 \pi)$ for the magnetic dipoles of magnitude $\mu$. Here $\epsilon_0$ and $\mu_0$ give, respectively, the permittivity and permeability of vacuum.

Table~\ref{tab:species} illustrates the strength of the dipole-dipole interaction parameters that can be achieved with several ultracold species currently available in the laboratory. For the molecules, the dipoles can be conveniently polarized using a microwave field coupling the two lowest rotational states, $J=0$ and $J=1$~\cite{LemeshkoPRL12, YanNature13}. In this case, due to the contributing Clebsch-Gordan coefficients, the resulting magnitude of the dipole-dipole interactions of Table~\ref{tab:species} needs to be divided by a factor of six. For a static field with a magnitude $E$, the ratio, $d/d_\text{mol}$,  of the induced dipole moment to the molecular one can be estimated in the strong-field limit as $d/d_\text{mol} = \left[ 1 - (2 d_\text{mol} E /B)^{-1/2} \right]$~\cite{LemMusKaisFriLong}, where $B$ is the molecular rotational constant. Experimentally feasible fields thus allow to achieve $d/d_\text{mol}\sim 0.7-0.8$~\cite{LemKreDoyKais13}, which results in the reduction of the dipole-dipole interaction matrix elements by a factor of two compared to the values listed in Table~\ref{tab:species}. We see that for typical values of lattice hoppings, $t \sim 10 - 10^3$~Hz, values of $c_d \gtrsim t$ are achievable for dipolar molecules; furthermore, the regime of substantial  magnitudes of $c_d$ can be accessible with magnetic atoms. We note, moreover, that recently created ultracold Er$_2$ molecules~\cite{Frisch15}, if prepared in their  fermionic incarnation, can in principle allow for $c_d = 135.2$~Hz on a lattice with $a=266$~nm. In experiment, the orientation of the dipoles with respect to the lattice, as given by the angles $(\theta, \phi)$ of Fig.~\ref{fig:setup}, can be controlled by tilting the polarization of the external microwave, electrostatic, or magnetic field~\cite{LemKreDoyKais13, BaranovCRev12, KreStwFrieColdMolecules, CarrNJP09}. 

The main idea of the dual boson method~\cite{Rubtsov12} is to separate the local and nonlocal physics from each other. As in DMFT, the local physics is encapsulated in an auxilary single-site impurity problem. Since the impurity problem possesses only a few degrees of freedom, it can be solved numerically exactly even when the correlation effects are significant. 
The remaining nonlocal physics is decoupled using an exact transformation to the new -- \emph{dual} -- degrees of freedom.
The main correlation effects are taken into account at the level of the impurity problem, thus the dual degrees of freedom are only weakly correlated, and therefore can be treated using perturbation theory.
The lattice size enters  only the relatively simple dual part of the calculation, which allows to increase the system size at a reasonable computational cost. In turn, this allows to significantly increase the momentum resolution, which is required for studying long-range ordered phases. As an example, the method has  recently been applied to plasmons in two-dimensional strongly-correlated electron systems~\cite{vanLoon14}, where the long-range Coulomb interaction plays a crucial role~\cite{vanLoon14}.  More details about the computation scheme can be found in Refs.~\cite{sup, vanLoon14-2}.  

In the dual boson approach, a charge order instability is revealed as a divergence in the static density-density susceptibility, $X_{\qv}=\av{\rho\rho}_{\qv,\omega=0}$, with $\rho=n-\av{n}$. The latter can be obtained from the dual perturbation theory using an exact transformation~\cite{Rubtsov12}, and  satisfies the charge-conservation laws~\cite{Hafermann14-2}. The momentum $\qv$, in turn, characterizes the type of the emerging charge order. This way, the signatures of a phase transition are already visible as it is approached from an unordered phase. In experiment, this corresponds to real-space density correlations which emerge in the vicinity of the phase transition. In an ultracold setting, the density-density correlations can be detected using, e.g., Bragg scattering~\cite{Stamper-Kurn1999}, time-of-flight~\cite{JonaLasinioPRA13}, or noise correlation \cite{Altman2004,Folling2005} spectroscopy.

\begin{table}[t]
\caption{\label{tab:species} The dipole-dipole interaction strength, $c_d$ (in Hz), for selected fermionic species currently available in the laboratory. For molecules, the value of the molecular-frame dipole moment was used to evaluate $c_d$.}
\begin{tabular}{r l r r r}
\hline
\hline
 & & \multicolumn{3}{c}{$c_d$ (Hz)} \\
 & & $a=$ 1064 nm   & $a=$ 532 nm  & $a=$ 266 nm \\
\hline
$^{23}$Na$^{40}$K&\cite{ParkPRL15, Park15b} & 926 & 7.4$\times10^3$ & 59.3$\times10^3$   \\
$^{40}$K$^{87}$Rb&\cite{NiNature10, OspelkausScience10} & 40 & 321 & 2.6$\times10^3$   \\
$^{161}$Dy&\cite{LuPRL12} & 1 & 8.6 & 68.7  \\
$^{167}$Er&\cite{AikawaPRL14, AikawaScience14} & 0.5 & 4.2 & 33.8   \\
$^{53}$Cr&\cite{NaylorPRA15} & 0.4 & 3.1 & 24.8   \\
\hline
\hline
\end{tabular}
\end{table}

\begin{figure}
 \includegraphics[width=\linewidth]{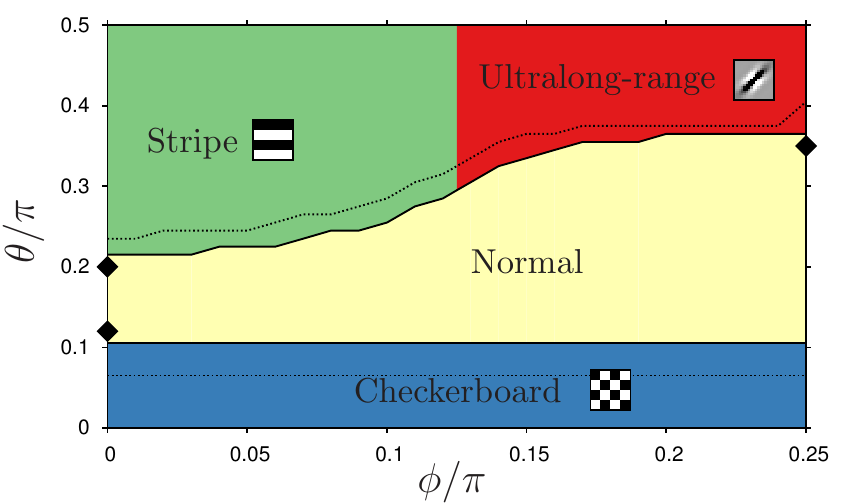}
 \caption{(Color online) Phase diagram as a function of the dipole orientation at $U=4t$, $c_d = 2t$, and filling $\av{n} \approx 0.9$. The dotted lines show the phase boundaries at the reduced dipolar coupling, $c_d=1.8 t$. The black diamonds show the angles selected for Fig.~\ref{fig:xlat}.   
 }
 \label{fig:pd1}
\end{figure}

We now evaluate the phase diagram of the DFH Hamiltonian (\ref{eq:hmlt}) with  a repulsive on-site interaction $U=4t$, the dipole-dipole interaction strength $c_d = 2t$, and filling of $\av{n} \approx 0.9$ fermions per site, with equal populations of spin-up and spin-down states. In order to reproduce the conditions of ultracold experiments we set the temperature to $T = t/(4 k_B)$, where $k_B$ is Boltzmann's constant. As an example, for the hopping rate of $t = 2 \pi \hbar \times 1$~kHz this corresponds to $T = 12.5$~nK.  In order to reveal the effect of anisotropy on the many-body state of the system, we evaluate the phase diagram depending on the orientation of the dipoles with respect to the lattice plane. The resulting phase diagram is shown in Fig.~\ref{fig:pd1}. 

\begin{figure}
 \includegraphics{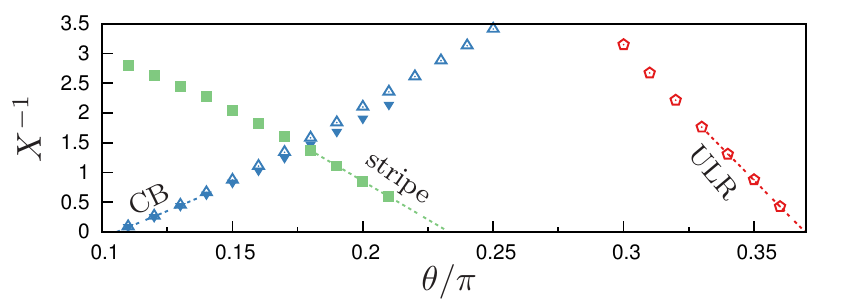}
 \caption{(Color online) Inverse charge susceptibility in the normal phase, for the same parameters  as in Fig.~\ref{fig:pd1}. The blue triangles show the divergence of the susceptibility at the checkerboard point, $\qv_{\text{CB}}=(\pi,\pi)$, as $\theta$ is lowered, both for $\phi=0$ (filled triangles) and $\phi=0.25\pi$ (empty triangles). The green squares show the divergence of the $\qv_{\text{stripe}}=(0,\pi)$ susceptibility as $\theta$ increases at $\phi=0$ and the red pentagons show the $\qv_{*}\approx (0.21\pi,-0.21\pi)$ susceptibility at $\phi=0.25\pi$. 
 The dashed lines show a linear extrapolation of the inverse susceptibility. }
 \label{fig:susc1}
\end{figure}

\begin{figure}
  \includegraphics[width=1.03\linewidth]{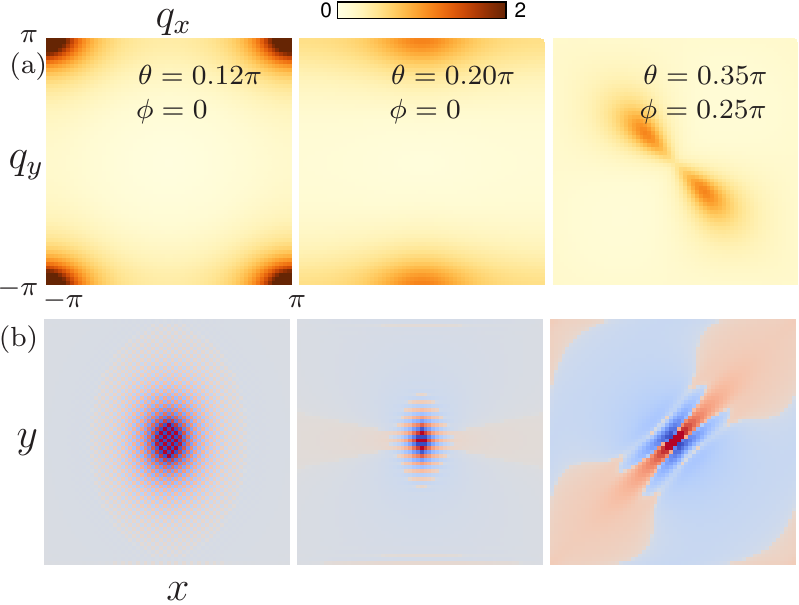}
  \caption{(Color online) (a) Momentum-space susceptibility at selected points of the phase diagram, Fig.~\ref{fig:pd1}. (b) The corresponding density correlation function in real-space: given a particle in the center of the figure, red indicates a higher probability to find a particle at $x,y$ and blue a lower probability. Every pixel corresponds to a lattice site. }
  \label{fig:xlat}
\end{figure}

Depending on the orientation of the dipoles, the system goes through different phases. In the middle of the phase diagram, there is a normal phase featuring no particular order (a ``metal'' or ``Fermi liquid'' state). At smaller values of $\theta$, which corresponds to the dipoles oriented nearly perpendicular to the lattice plane,
a transition to checkerboard order [with $\qv_{\text{CB}}=(\pi,\pi)$] occurs. Such a checkerboard phase has been previously observed in half-filled systems with nearest-neighbor interaction~\cite{Bari71,Vonsovsky79,Zhang89,Sun02,Ayral13,vanLoon14-2} and away from half-filling at $\theta=0$~\cite{Hofstetter11}.
In the normal phase, the signatures of the transition are already visible as the phase boundary of the checkerboard phase is approached: as shown in Fig.~\ref{fig:susc1}, the susceptibility at $\qv_{\text{CB}}=(\pi,\pi)$ diverges as $\theta$ is lowered. The susceptibility is shown both at $\phi=0$ (filled symbols) and $\phi=0.25\pi$ (empty symbols), the checkerboard susceptibility depends on $\phi$  only weakly.
The left panels of Fig.~\ref{fig:xlat} show the charge susceptibility close to the phase boundary, at $\theta=0.12\pi$, $\phi=0$. 
The momentum-space susceptibility has a clear maximum at $\qv_{\text{CB}}=(\pi,\pi)$.
The sign of the real-space charge correlation function, shown in panel (b), features a checkerboard pattern. 

At larger $\theta$, when dipoles are oriented nearly in the lattice plane, other charge-ordered phases occur in Fig.~\ref{fig:pd1}. At small $\phi$ (dipoles pointing along $\hat{x}$), there is a horizontally striped phase, whose charge order is given by the momentum $\qv_{\text{stripe}}=(0,\pi)$. Again, the tendency towards a diverging susceptibility is already visible in the normal phase, see Fig.~\ref{fig:susc1}. The real and momentum space susceptibility close to the phase boundary, at $\theta=0.20\pi$, $\phi=0$, shows a stripe pattern at small $x$. As the phase boundary is approached, the striped order takes over also at longer wavelengths (see Supplemental Material). A similar striped phase has been predicted before~\cite{Bhongale12}. 

 \begin{figure}
  \includegraphics[width=1.03\linewidth]{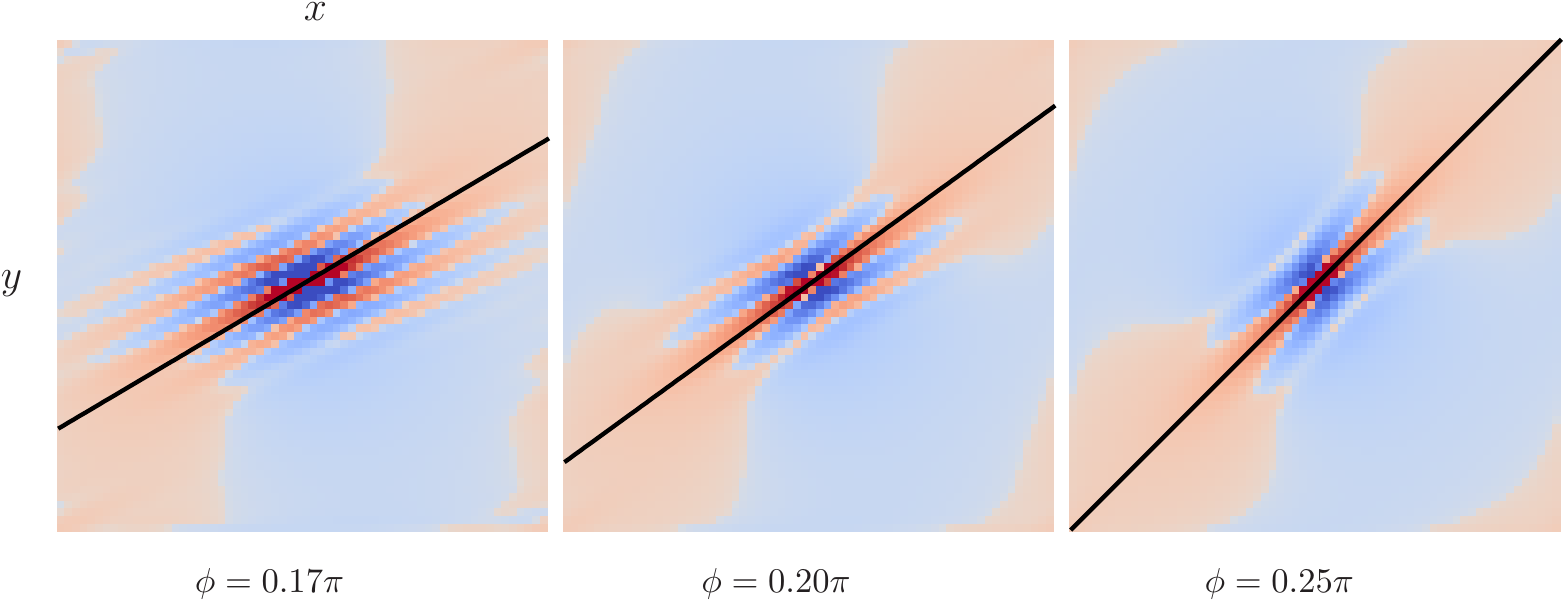}
  \caption{(Color online) 
  Real space density correlation function as in Fig.~\ref{fig:xlat}~(b), at fixed $\theta=0.35\pi$, for three values of $\phi$. The areas of high and low density follow the angle $\phi$ (black line).
  }
  \label{fig:xlat_phi}
 \end{figure}
 
Finally, when both $\theta$ and $\phi$ are large, corresponding to the dipoles oriented along the $xy$-diagonal, we find a novel phase possessing an ultralong-range order. The right-hand side of Fig.~\ref{fig:xlat} shows the susceptibility close to the transition towards this ordered phase. Continuous areas of high and low density extend over a large number of lattice sites. The maximum of the momentum-space susceptibility also occurs at smaller $\qv$ (longer wavelength).
Unlike in the other two ordered phases, here the shape of the real-space correlation function strongly depends on the angle $\phi$, with the areas of high and low density rotating along with the dipole orientation in the plane.
The ordering vector $\qv_{*}$ depends on $\phi$ and evolves from $\qv_{*}\approx\qv_{\text{stripe}}=(0,-\pi)$ close to the border with the striped phase to $\qv_{*}\approx (0.21\pi,-0.21\pi)$ at $\phi=0.25\pi$.
Fig.~\ref{fig:xlat_phi} shows that the high density parts of the real-space susceptibility follow the dipole angle $\phi$. The angular dependence of the density correlation function reflects the anisotropy of the dipole-dipole interaction, as given by the second spherical harmonic.

The spatial symmetry of the system helps to get insight into the different orderings. When the dipoles are oriented in the $z$-direction [$\theta=0$], the system possesses mirror symmetry with respect to the $x$- and $y$-axis and the $xy$-diagonal, which coincides with the symmetry of the checkerboard pattern.
For dipoles oriented along the $x$-axis [$\theta=\pi/2$, $\phi=0$], the mirror symmetry along the diagonal is broken by the dipole orientation. Indeed, the horizontally-striped phase has the mirror symmetries only with respect to the $x$- and $y$-axes.
Finally, for dipoles oriented diagonally in-plane [$\theta=\pi/2$, $\phi=\pi/4$], the system has mirror symmetry around the two in-plane diagonals -- exactly as the resulting ultralong-range charge order.
 
Let us discuss the stability of the observed phase diagram with respect to changing the parameters of the Hamiltonian (\ref{eq:hmlt}). The ordered phases originate from the interplay between the contact and the long-range interaction terms, therefore the phase boundaries shift depending on the value of $c_{d}$. This point is illustrated in Fig.~\ref{fig:pd1} by the dotted lines, corresponding to the phase boundaries at a reduced dipolar interaction, $c_d=1.8t$. The structure of the phase diagram, however, stays qualitatively similar. In addition, we have studied the phase diagram as a function of the filling $\av{n}$. At $\av{n}\approx 0.8$, the phase diagram is qualitatively similar to Fig.~\ref{fig:pd1}, while at a further reduced value  of  $\av{n}\approx 0.49$ the checkerboard instability disappears at small $\theta$. In Fig.~\ref{fig:pd1}, the normal, striped and ultralong-range phases meet near $\phi=0.12\pi$. The location of the triple point depends on the density $\av{n}$ and the dipole strength $c_d$, however we found it always to be close to $\phi=0.12\pi$. This indicates that in experiment small density fluctuations due to a harmonic trap are unlikely to qualitatively change the observed order. Furthermore, ultracold gases in uniform potentials with constant density have been created in laboratory~\cite{GauntPRL13}. 

In order to study the importance of the long-range character of the dipole-dipole interactions, we have performed simulations with the interaction $V^{d}_{jk}$ restricted to the nearest neighbours. In such a case, the susceptibility near the boundaries with the checkerboard and striped phases looks qualitatively similar. On the other hand, the dumbbell-shaped density correlation function that occurred in Fig.~\ref{fig:xlat} at $\theta=0.35\pi$, $\phi=0.25\pi$, disappears completely. Therefore, we  conclude that while the checkerboard and striped phases are mainly driven by the nearest-neighbor interaction,  the long-range couplings are essential for the ultralong-range ordered phase to form.

Large-scale simulations appear to be vital in order to capture and describe the ultralong-range order. As illustrated in the Supplemental Material~\cite{sup}, in simulations involving smaller lattice sizes finite-size effects can significantly affect the results.

Thus, we have demonstrated the occurrence of a novel, ultralong-range-ordered phase in the Fermi-Hubbard model with dipole-dipole interactions. This reveals the importance of taking into account large system sizes while dealing with the species featuring anisotropic long-range interactions. The novel phase should be within reach for current experiments with ultracold polar molecules and magnetic atoms. The formation of the ultralong-range order might be a general phenomenon for systems with interactions beyond the nearest neighbor, e.g. those involving quadrupoles~\cite{BhongalePRL13, LarzPRA13, LahrzNJP15} and oscillating light-induced dipoles~\cite{LemeshkoPRA11Optical, LemFri11OpticalLong}.

\begin{acknowledgments}

We are grateful to Sebastian Will for fruitful discussions. We used an extended version of the CT-HYB solver of Ref.~\cite{Hafermann13}. The solver and dual boson implementation are based on the ALPS libaries~\cite{ALPS2}. The work is supported by European Research Council (ERC) Advanced Grant No.~338957 FEMTO/NANO.
\end{acknowledgments}

\appendix

\section{Finite-size effects}

\begin{figure}[h]
  \includegraphics[width=\linewidth]{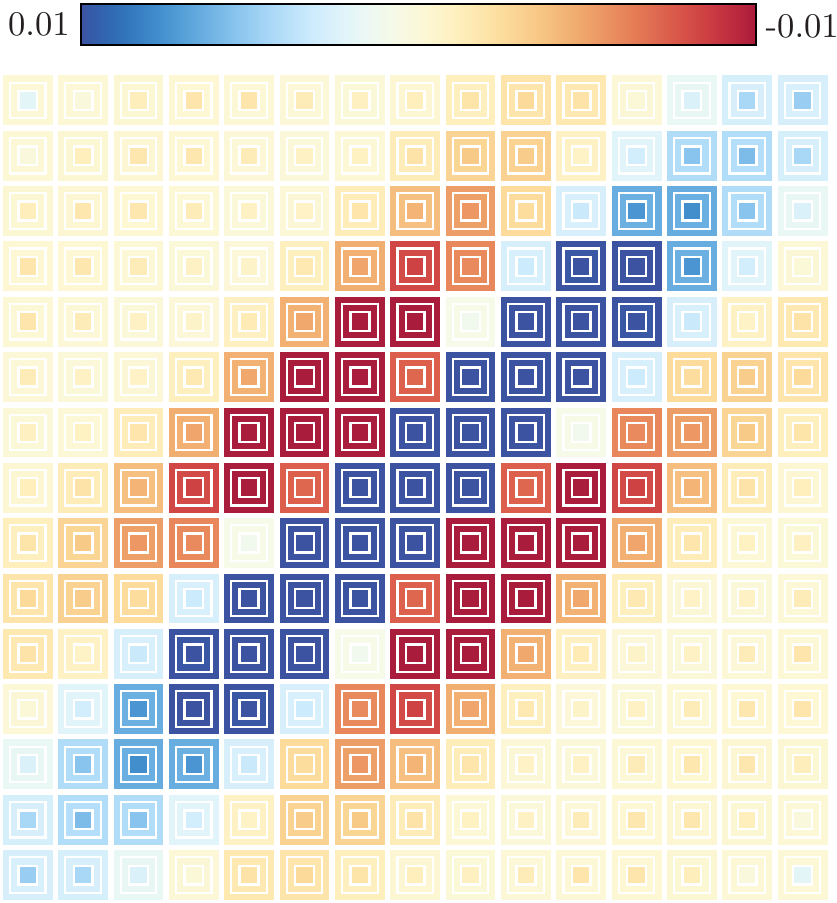}
  \includegraphics[width=\linewidth]{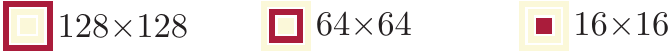}
  \caption{(Color online)
  The real-space correlation function at $\theta=0.35\pi$, $\phi=0.25\pi$, for three different lattice sizes.
  Values are rounded to the range of $[-0.01,0.01]$. The 64$\times$64 and 128$\times128$ results are very similar;  the 16$\times$16 result, on the other hand, reveals finite size effects.
  }
  \label{fig:finitesize}
 \end{figure}

The simulations were performed on a $64 \times 64$ square lattice  with periodic boundary conditions. Selected points in the phase diagram were also calculated on a $128 \times 128$ lattice in order to rule out the finite-size effects. For comparison, we have also done calculations on a $16 \times 16$ lattice. There, finite size effects are clearly visible and the momentum resolution is insufficient to accurately determine the ultralong-range order. The comparison of the results for different lattice sizes is shown in Fig.~\ref{fig:finitesize}.

\section{Density}

The dual boson approach works in the grand canonical ensemble, therefore all results are obtained at a fixed chemical potential. We fixed the chemical potential after subtracting the Hartree contribution of the nonlocal interaction. 
We note that while scanning over the dipole orientation, as in Fig.~2 of the main text, the density cannot be held exactly constant since the dipolar interaction also gives beyond-Hartree contributions to the density. However, the difference in density between different points in the phase diagram is on the order of $\Delta\! \av{n}\approx 0.01$ and is therefore negligible.

\section{Computational scheme}

In the calculation of the susceptibility, we first calculated the susceptibility of the dual particles by summing all particle-hole ladder diagrams. In this way, we treated repeated scattering processes to all orders and the resulting susceptibility satisfies the charge-conservation requirements~\cite{Rubtsov12,Hafermann14-2}.
 
Charge-order transitions can manifest themselves in two ways in our calculations. As explained in the main text, a divergence in the dual boson charge susceptibility is a sign of an ordered phase.
However, sometimes the EDMFT self-consistency (which is done prior to the dual boson part) already shows a diverging susceptibility. In such a case, converging to EDMFT self-consistently is difficult. Previous work on the checkerboard charge-ordering transitions in the extended Hubbard model showed that the dual boson transition occurs already at a smaller interaction strength compared to the EDMFT transition~\cite{vanLoon14-2}. However, the region between the dual boson ordering and EDMFT ordering can be small. 
For the results of Fig.~2 of the main text, EDMFT converged in the checkerboard region, and the phase transition became visible when the nonlocal dual boson corrections were added to the susceptibility. In the striped and ultralong-range ordered phase at large $\theta$, on the other hand, the EDMFT part of the calculation already suffers from divergencies. Here, following the susceptibility from the normal phase is the most stable way to determine the type of order. The location of the triple point of the normal phase, where it meets with the striped and ultra-long-range order, can be determined by approaching it from the normal phase. However, using our method, it is challenging to predict how the the phase boundary between the striped and ultralong-range ordered phases behaves for angles $\theta$ above the triple point.
 
To study the properties of the normal phase (without order), we have looked at the single-particle Green's function in imaginary time, $G(\tau=\beta/2)$, which gives an estimate for the amount of spectral weight around the Fermi energy. This procedure was also used in Ref.~\onlinecite{vanLoon14-2}. Additionally, the compressibility $d\av{n}/d\mu=X_{q=0,\omega=0}$ can be obtained directly from the susceptibility. Both quantities indicate that the normal phase is metallic.

\section{Approaching the transition}

\begin{figure*}[b]
  \includegraphics[width=\linewidth]{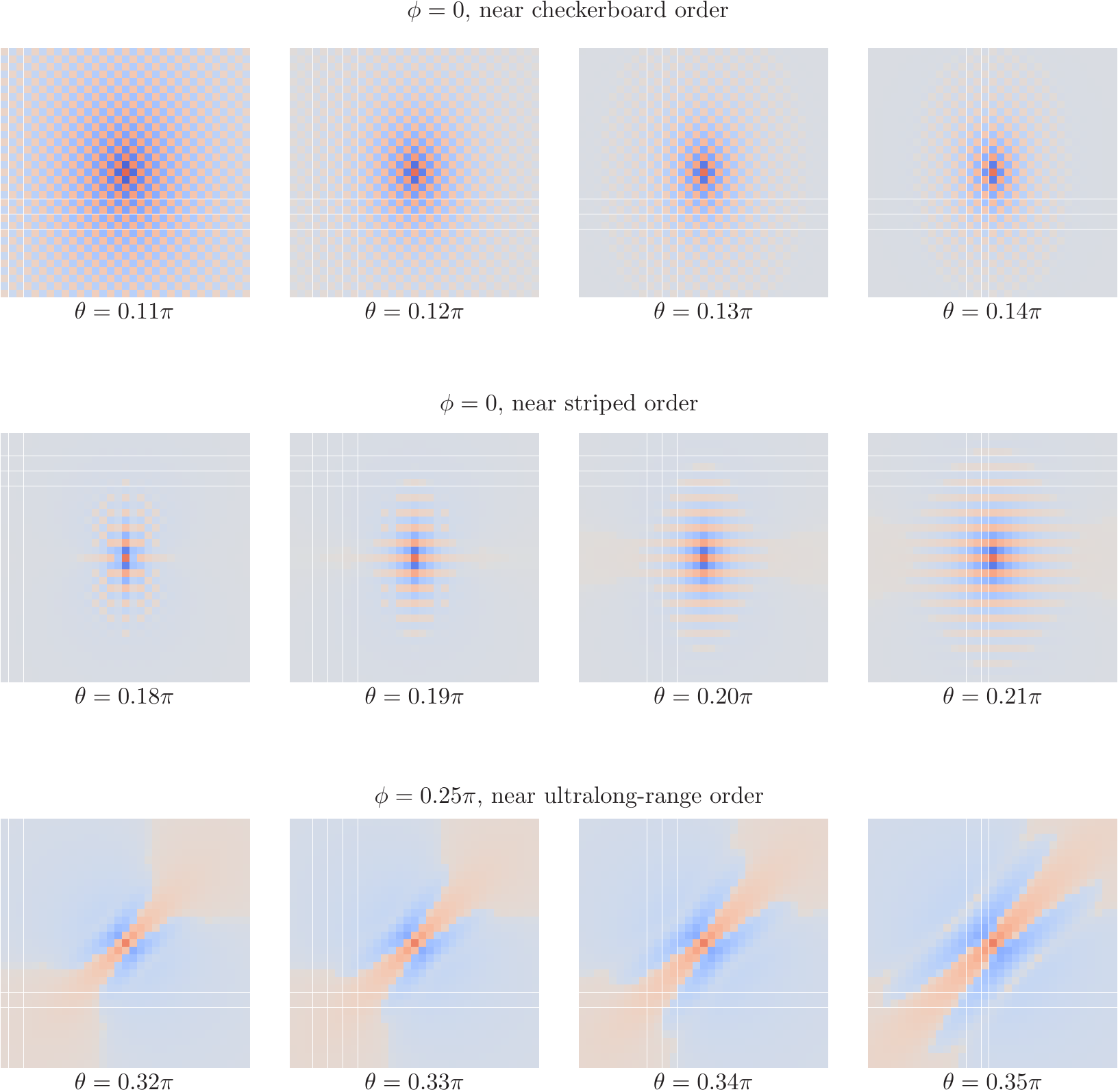}
  \caption{
  The real-space correlation function close to the phase transitions.
  \label{fig:approach}
  }
 \end{figure*}

In Fig.~\ref{fig:approach}, we show the real-space correlation function as the ordered phases are approached. These figures are zoomed in, compared to the figures of the main text, only the sites close to the origin are shown.

\bibliography{DipolarFermions}

\end{document}